\documentclass[hidelinks]{article}
\usepackage{arxiv}

\usepackage{amsthm}

\usepackage{graphicx} 
\usepackage{caption}
\usepackage{subcaption}
\usepackage{enumerate}
\usepackage{multicol}
\usepackage{listings}
\usepackage{color}
\usepackage[normalem]{ulem}
\definecolor{light-gray}{gray}{0.95}

\title{LMN: A Tool for Generating Machine Enforceable Policies from Natural Language Access Control Rules using LLMs 
}

\author{%
  Pratik Sonune\\
  Indian Institute of Technology Kharagpur, India\\
  \texttt{pratiksonune28@gmail.com} 
  \And
 Ritwik Rai\\
  Indian Institute of Technology Kharagpur, India\\
  \texttt{rajveenrai36@gmail.com} 
  \And
  Shamik~Sural \\
  Indian Institute of Technology Kharagpur, India\\
  \texttt{shamik@cse.ac.in} 
  \AND
  Vijayalakshmi Atluri \\
  Rutgers University, Newark, USA\\
  \texttt{atluri@rutgers.edu } 
  \And
  Ashish Kundu \\
  CISCO Research, USA\\
  \texttt{ashkundu@cisco.com} \\
}

\date{}

\renewcommand{\headeright}{}
\renewcommand{\undertitle}{}
\renewcommand{\shorttitle}{Generating Machine Enforceable Policies from Natural Language Access Control Rules using LLMs}

\begin{document}
\maketitle
\begin{abstract}
Organizations often lay down rules or guidelines called Natural Language Access Control Policies (NLACPs) for specifying who gets access to which information and when. However, these cannot be directly used in a target access control model like Attribute-based Access Control (ABAC). Manually translating the NLACP rules into Machine Enforceable Security Policies (MESPs) is both time consuming and resource intensive, rendering it infeasible especially for large organizations. Automated machine translation workflows, on the other hand, require information security officers to be adept at using such processes. To effectively address this problem, we have developed a free web-based publicly accessible tool called LMN (\underline{L}LMs for generating \underline{M}ESPs from \underline{N}LACPs) that takes an NLACP as input and converts it into a corresponding MESP. Internally, LMN uses the GPT 3.5 API calls and an appropriately chosen prompt. Extensive experiments with different prompts and performance metrics firmly establish the usefulness of LMN.
\end{abstract}

\keywords{Attribute-based Access Control, Natural Language Access Control Policy, Machine Enforceable Security Policy, Large Language Models, Prompt Engineering, Web Application
}

\section{Introduction}
\label{sec:intro}

Access control is a fundamental security requirement in any organization for ensuring that only authorized users can access certain information or resources under specific conditions. While enforcement needs to be done in computer systems, access control policies are typically decided by the higher management. For example, in a university system, the Department Chair, Dean and the Provost may take a decision on who can access which object (like Conference room printers, Graduate studies applications, Faculty tenure support letters, etc.) at the Department, School and University level, respectively. Such decisions are often noted down as meeting minutes, email exchanges, or other forms of documentation in a natural language like English (hereinafter referred to as Natural Language Access Control Policies, i.e., NLACPs). For information system level implementation of such decisions, System Security Officers (SSOs) must translate the NLACPs into  Machine Enforceable Security Policies (MESPs) using a target access control model like Role-based Access Control (RBAC) or Attribute-based Access Control (ABAC). 

It is apparent that manual conversion of NLACPs into MESPs not only demands time and resource, it is also error prone, especially for large organizations with dynamically changing policies. While recent advancements in Natural Language Processing (NLP) techniques, specifically Large Language Models (LLMs), can be made use of, the SSOs need to be aware of the ever expanding functions and capabilities of these tools. Although SSOs are highly skilled with information security standards and procedures, expecting them to know the nuances of various LLMs and their new releases is an additional burden. 

With an aim to addressing this particular problem, we have built and deployed a web-based tool called LMN (\underline{L}LMs for generating \underline{M}ESPs from \underline{N}LACPs) and made it publicly available for free usage. It takes natural language access control policies and optionally a list of attributes as input, and produces ABAC MESPs as output by invoking the APIs of the GPT 3.5 LLM. We also carried out several experiments with different types of prompts and extracted a number of NLP performance metrics. The results look promising.

\section{Preliminaries}
\label{sec:prelim}

In this section, we first briefly describe the components of Attribute-based Access Control and then introduce some of the basic concepts behind LLMs, including GPT which has been used in our LMN tool. 

\subsection{Attribute Based Access Control}
\label{subsec:abac}
ABAC is a dynamic and adaptive model of access control that has won much attention in modern information security systems. It is based on the notion of attributes which are characteristics of entities. There are primarily three types of entities, namely users or subjects (U), resources or objects (O) and environmental conditions (E). U is the set of entities that request for access to resources, O is the set of resources to be protected from unauthorized access while E denotes the set of possible Contexts in which access requests are made. Each of these entities is characterized by attributes, i.e., user attributes \(a^u\), object attributes \(a^o\) and environmental attributes \(a^e\). Activities that a given user performs on a particular object in a specific environmental condition is called an operation \(op\). For example, in a university system, the user attributes could be \textit{role, department}, the object attributes could be \textit{doc\_type, creation\_date, owner}, while the environmental attributes could be \textit{location, day, time}.
Another important component of the ABAC model is the \textit{policy}. A Policy (P) in ABAC is a set of authorization rules that governs who can access what object under which environmental conditions.

Each rule $r_i \in P$ is of the form $C_i^u \wedge C_i^o \wedge C_i^e \wedge op_i$. Here $C_i^u , C_i^o  and  C_i^e$ represent user condition, object condition and environmental condition respectively. These conditions can be expressed as follows: 
$$a_i^u = v_1 \wedge a_2^u = v_2 , \dots \wedge a_m^u = v_m \Rightarrow C_i^u$$
$$a_i^o = v_1 \wedge a_2^o = v_2 , \dots  \wedge a_n^o = v_n \Rightarrow C_i^o$$
$$a_i^e = v_1 \wedge a_2^e = v_2 , \dots  \wedge a_k^e = v_k \Rightarrow C_i^e$$

It is evident from the above discussion that for an ABAC system to be functional, the set of rules in the ABAC policy $P$ of the organization needs to be first constructed and then used for enforcement when an access request arrives. 

\subsection{Large Language Models and GPT}
\label{subsec:llm}

Research on NLP has long been heavily dependent on the availability of labeled data and access to powerful computing systems (typically with GPU) for large model training. Further, the primary goal was to interpret and potentially translate text from one language to another. In contrast, recent years saw an enormous growth in the availability and popularity of Large Language Models, which are models with billions of parameters that have been pre-trained with a huge corpus and can be employed in a generative setting. Examples of LLMs include GPT, LLama, OPT, Gemini, etc.  

While any of the above-mentioned LLMs could have been used for building our MESP generation tool LMN, we particularly employ the Generative Pre-trained Transformer (GPT) model\footnote{https://platform.openai.com/docs/models/gpt-3-5}. 
The transformer architecture in GPT makes use of self-attention, which further allows the network to consider all the contexts of a whole phrase for generation of the next word, thus improving the ability of the model to comprehend and generate sentences in a given language. The output text is generated by a decoder based on the input representation. Any such Large Language Model including GPT requires an input along with a prompt that guides the LLM to generate the type of output the user would be interested in.

\begin{figure*}[t]
    \centering
\includegraphics[width=1.0\textwidth]{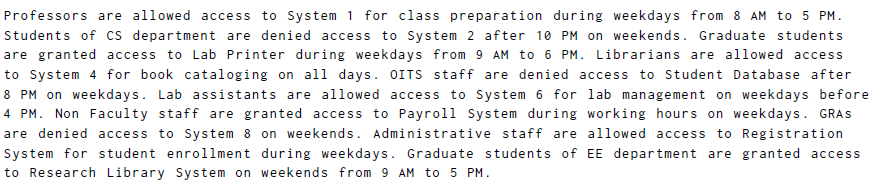}
        \caption{Example Input Natural Language Access Control Policy}
    \label{fig:nlacp_example}
\end{figure*}

\begin{figure*}[t]
    \centering
\includegraphics[width=0.7\textwidth]{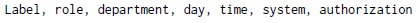}
        \caption{Example Input Attribute List}
    \label{fig:attrib_example}
\end{figure*}

\begin{figure*}[t]
    \centering
\includegraphics[width=1.0\textwidth]{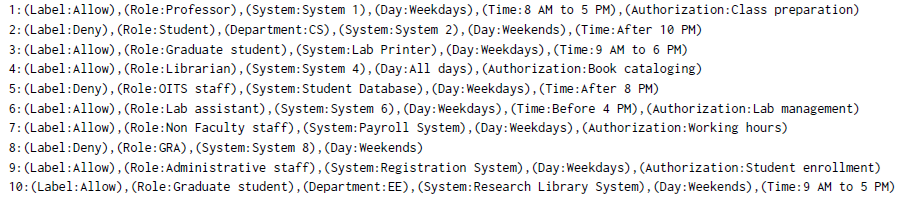}
        \caption{LMN2 Output}
    \label{fig:lmn2_output_example}
\end{figure*}

\begin{figure*}[t]
    \centering
\includegraphics[width=1.0\textwidth]{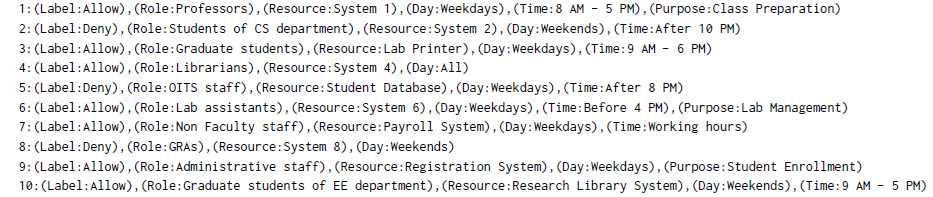}
        \caption{LMN1 Output}
    \label{fig:lmn1_output_example}
\end{figure*}

\section{Functionality and Design of LMN}
\label{sec:LMNfunctionality}

The motivation behind building the LMN tool is to automate the process of conversion of NLACPs into MESPs, and make it available to the community as a freely usable web-based tool. The functionality and design details of LMN are described in the following sub-sections.

\subsection{Input Requirements and Output}
\label{subsec:input&output}

Based on the level of information available with the SSO of an organization, LMN supports two forms of inputs. If the SSO only has the NLACP, a single text file (We refer to it as the NLACP file) needs to be used in LMN for generating the MESP. On the other hand, if the SSO has additional information about the potential list of attributes, the same may also be used (We refer to it as the Attributes file) along with the NLACP. In the rest of our discussions, we call these as the one-input version (or LMN1) and the two-input version (or LMN2), respectively. In either case, the SSOs do not need to provide any other input.
In both cases, the Machine Enforceable Security Policy is generated as output, which is referred to as the MESP file.

An example NLACP and an attributes file are shown in Figures \ref{fig:nlacp_example} and \ref{fig:attrib_example}, respectively. The corresponding MESP output of LMN2 is shown in Figure \ref{fig:lmn2_output_example} and that of LMN1 is shown in Figure \ref{fig:lmn1_output_example}.

\subsection{Role of GPT and Processing Steps}
\label{subsec:roleofGPT}

LMN utilizes GPT 3.5 of OpenAI for transforming NLACPs into MESPs (Note that, Other versions of GPT would also work seamlessly). The model interacts with the input data through API calls, and the output is generated by processing the prompts designed for the respective versions LMN1 and LMN2. API integration with GPT is done using a secure API key, ensuring data confidentiality. The raw GPT output is further refined for accuracy. By using an LLM like GPT 3.5, we minimize human intervention in access control policy conversion, thereby reducing possibilities of errors and accelerating the process. This ensures that the generated policies are consistent and accurate compared to manually converting natural language policies.
The main processing steps of LMN include Input Handling, GPT 3.5 Interaction and Post-Processing.\\
\textbf{Input Handling}:
A user uploads the required input files (\texttt{NLACP.txt} and optionally \texttt{attributes.txt}) via a web-based application interface. Depending on the selected version (LMN1 or LMN2), the uploaded files are processed.\\
\textbf{GPT 3.5 Interaction}:
A prompt is sent from the web application to GPT 3.5 through an API call, specifying either to generate ABAC rules based on the given attributes (LMN2) or to dynamically extract the attributes first and then generate ABAC rules (LMN1). The model processes the inputs and returns the generated ABAC rules in a structured format as MESP. \\
\textbf{Post-Processing}:
The output from GPT 3.5 is saved in two files, namely, \texttt{MESP.txt} - containing the generated ABAC rules and \texttt{gpt\_attribute.txt} - listing the attributes used. The files are compressed into a ZIP archive and downloaded to the user's machine.

\subsection{Development of Prompts}
\label{subsec:promptengg}

Although LLMs and generative AI have made significant progress in recent years, it has become equally crucial to learn how to effectively utilize the power of these models and do so efficiently. Prompts are texts or instructions that are provided to any generative AI model for extracting desired information from them. Appropriate prompting is important as ambiguous and unclear prompts lead to poor response. Prompt engineering \cite{chen2024unleashingpotentialpromptengineering} deals with developing effective and optimized prompts when interacting with LLMs. 

While building the LMN tool, different prompts were designed to achieve two types of policy generation as mentioned below. For LMN2, a prompt was designed to provide GPT 3.5 with both the NLACP policies and the associated attributes. This way, the model has sufficient information about the context and attributes, leading to more accurate ABAC rule generation. On the other hand, for LMN1, the prompt directs GPT 3.5 to analyze the provided natural language policies and extract the relevant attributes on its own. This requires a well-crafted prompt to guide the model in identifying attribute-value pairs without any additional information.

Effective prompt engineering while designing LMN ensured that the model generates accurate access control rules, reduces ambiguities, and produces consistent results. During the development phase of the tool, different prompts were iteratively tested and refined to improve the output quality. 
For example, prompts were designed to include specific instructions as follows:

\begin{itemize}
\item \textbf{Clear Description of the Desired Output}: The prompt included a clear description of the expected output format, such as a list of ABAC rules with attribute-value pairs.
\item \textbf{Attribute Extraction Instructions}: In the case of LMN1, the prompt asked the model to identify attributes like role, resource, and conditions dynamically from the given text. This approach leveraged the model’s understanding of natural language to accurately extract necessary details.
\item \textbf{Example Formatting}: To aid the model, example formatting of the output was included within the prompt, showing a structured ABAC rule format like: \texttt{(Label: Allow), (Role: User), (Resource: System)}. This helped in maintaining a consistent output structure.
\end{itemize}

\section{LMN Implementation Details}
\label{sec:impl}
\begin{figure*}[ht!]
    \centering
\includegraphics[width=1.0\textwidth]{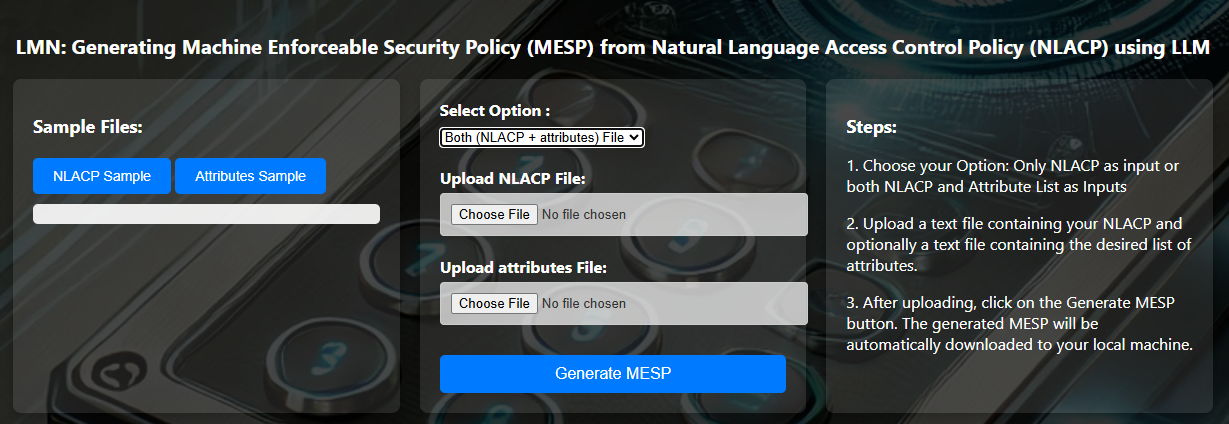}
        \caption{Web Interface of LMN}
    \label{fig:screenshot}
\end{figure*}

We have implemented LMN as a web application\footnote{https://llm.abac.iitkgp.ac.in/} hosted on a server and accessible from anywhere free of cost. In this section, we provide the intricate details of its implementation that shows the challenging aspects of building the tool. A screenshot of the application is shown in Figure \ref{fig:screenshot}.

\subsection{Flask Framework}
\label{subec:flaskframework}
LMN leverages the Flask framework of Python for creating a user-friendly interface. Flask was chosen due to its flexibility, ease of use, and the ability to support RESTful API integration seamlessly. It provides the following advantages.

\begin{itemize}
    \item \textbf{Simplicity and Flexibility}: Flask’s minimalist design allows developers to create simple yet powerful web applications without a lot of configuration overhead.
    \item \textbf{Integration with External APIs}: Flask easily integrates with the OpenAI API for calling GPT 3.5 to generate the MESPs. This API integration is crucial for sending prompts and receiving structured output from the model.
    \item \textbf{File Handling Capabilities}: Flask's native support for file uploads and routing makes it well-suited for creating an application that handles text files for input. Users can easily upload the NLACP and the Attributes files for processing.
    \item \textbf{Rapid Development}: Flask allows for rapid prototyping and development, which was essential in testing different versions (LMN1 and LMN2) and refining the prompts used for LLM interaction. 
\end{itemize}

In LMN, the Flask framework facilitates user interaction by providing an interface where users can upload files and receive the generated machine-enforceable policies as a download, simplifying the entire process of policy transformation. 

\subsection{User Interaction Flow}
\label{subsec:userinteractionflow}
The implemented web-based LMN tool involves the following steps for user interaction:
\begin{enumerate}[i]
    \item User opens the web interface and selects the version of LMN to process (LMN1 or LMN2).
    \item The user uploads the required input files (\texttt{NLACP.txt} and optionally \texttt{attributes.txt}) and clicks on the \textit{Generate MESP} button.
    \item Flask handles the API call to GPT 3.5, sending the appropriate prompt for either dynamic attribute extraction or structured attribute input.
    \item The generated ABAC rules and attribute list are saved, and the user is provided with an auto-downloaded ZIP file containing the results.
\end{enumerate}

This approach ensures that organizations can easily convert their natural language access control rules into machine readable policies without requiring specialized knowledge about natural language processing in general and LLMs in particular.

\begin{table}[t]
    \centering
    \caption{Prompt Analysis for LMN2}
    \begin{tabular}{|c|c|c|c|c|c|c|}
        \hline
        & \textbf{P1} & \textbf{P2} & \textbf{P3} & \textbf{P4} & \textbf{P5} & \textbf{P6} \\
        \hline
        \textbf{Role} & 20 & 20 & 20 & 20 & 0 & 0 \\
        \hline
        \textbf{Department} & 20 & 20 & 20 & 0 & 0 & 0 \\
        \hline
        \textbf{System} & 20 & 20 & 20 & 20 & 0 & 20 \\
        \hline
        \textbf{Time} & 20 & 16 & 20 & 20 & 0 & 20 \\
        \hline
        \textbf{Day} & 20 & 20 & 20 & 20 & 0 & 0 \\
        \hline
        \textbf{Label} & 20 & 20 & 20 & 0 & 0 & 20 \\
        \hline
    \end{tabular}
    \label{tab:NLACP2Analysis}
\end{table}

\begin{table}[t]
    \centering
    \caption{Prompt Analysis for LMN1}
    \begin{tabular}{|c|c|c|c|c|c|c|}
        \hline
        & \textbf{P1} & \textbf{P2} & \textbf{P3} & \textbf{P4} & \textbf{P5} & \textbf{P6} \\
        \hline
        \textbf{Role} & 20 & 20 & 20 & 0 & 0 & 0 \\
        \hline
        \textbf{Department} & 20 & 0 & 19 & 0 & 17 & 0 \\
        \hline
        \textbf{System} & 20 & 20 & 20 & 0 & 20 & 20 \\
        \hline
        \textbf{Time} & 20 & 19 & 20 & 0 & 20 & 20 \\
        \hline
        \textbf{Day} & 20 & 19 & 20 & 0 & 0 & 0 \\
        \hline
        \textbf{Label} & 20 & 20 & 20 & 0 & 20 & 0 \\
        \hline
    \end{tabular}
    \label{tab:NLACP1Analysis}
\end{table}
\section{Experimental Results}
\label{sec:experiment}

As mentioned in the earlier sections, accuracy and effectiveness of a tool like LMN depends on the quality of prompts used. We, therefore, initially experimented with several prompts in the prompt engineering phase. These results are presented in the first sub-section below. Accuracy and efficiency results are then studied in the subsequent sub-sections.

\subsection{Prompt Engineering Results}
\label{subsec:promptresults}
We designed and tested six different prompts for determining their effectiveness in converting NLACPs into structured ABAC rules. 
The prompts were used to evaluate how well the model extracted the following attributes: Role, Department, System, Time, Day, and Label. The effectiveness of each prompt was quantified based on the number of times each attribute was correctly extracted, providing a comprehensive analysis of which prompt yielded the best results. Six prompts (P1 to P6) were used in the study, the details of which are given in Appendix \ref{apndx_sec_prompts}. Each prompt is designed with different levels of complexity and structure for both LMN1 and LMN2:

Table \ref{tab:NLACP2Analysis} presents the results for LMN2, where
the values represent the number of correct extractions for each attribute over twenty sample inputs. Likewise, Table \ref{tab:NLACP1Analysis} presents the results for LMN1.
From the tables, the following observations can be made. Prompts 1 and 3 generally performed the best, providing accurate extraction across all attributes for both LMN1 and LMN2. Prompts 4, 5 and 6 were the worst performers, with several values either incorrect or missing. Using extremely simple or overly complex prompts led to lower attribute extraction accuracy. A balanced prompt with a clear and structured approach yielded the best results.
This analysis helped in understanding the impact of prompt engineering on the quality of generated ABAC rules. The insights gained through this evaluation were instrumental in optimizing the process of transforming natural language policies into structured ABAC rules.

\begin{figure}[t]
\centering
\includegraphics[width=0.6\textwidth]{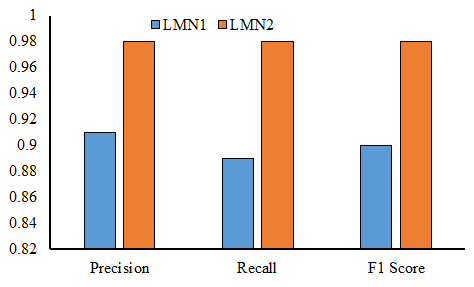}
        \label{fig:bert}
        \caption{BERT Scores}
    \label{fig:bertscores}
\end{figure}

\begin{figure*}[t]
    \begin{subfigure}[b]{0.5\textwidth}   \includegraphics[width=0.9\textwidth]{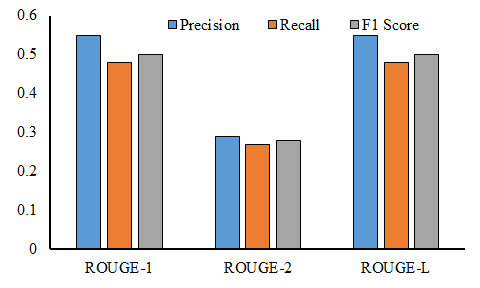}
        \caption{ROUGE Score Evaluation for LMN1}
        \label{fig:rouge1}
    \end{subfigure}
    \begin{subfigure}[b]{0.5\textwidth}
    \includegraphics[width=0.9\textwidth]{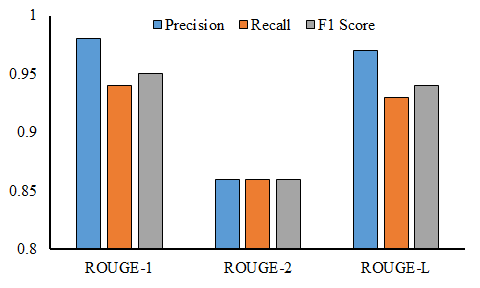}
        \caption{ROUGE Score Evaluation for LMN2}
        \label{fig:rouge2}
    \end{subfigure}
    \caption{ROUGE Scores}
        \label{fig:rougescores}
\end{figure*}

\begin{figure}[t]
    \centering
    \includegraphics[width=0.6\textwidth]{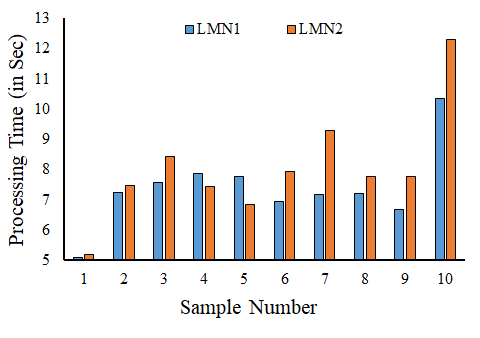}
    \caption{LMN1 and LMN2 Processing Time for 10 samples}
        \label{fig:time}
\end{figure}

\subsection{Accuracy and Efficiency Results}
\label{subsec:eval}

The following metrics were studied for evaluating the accuracy and efficiency of LMN. Prompt 1 as mentioned above was used for carrying out the experiments.

BERTScore evaluates the similarity between two texts by leveraging BERT (Bidirectional Encoder Representations from Transformers) embeddings \cite{devlin2019bertpretrainingdeepbidirectional}. Unlike traditional evaluation metrics, it captures semantic similarity and takes into account word meanings, synonyms and sentence structure. We used BERTScore to compare the ABAC rules generated by LMN with reference outputs for measuring how close the generated policies were to expected results. The evaluation was performed on ten sample inputs, and BERTScore was computed for both LMN1 and LMN2. The results are shown in Figure \ref{fig:bertscores}. It is observed that LMN2 generally performs better than LMN1 in terms of BERTScore, as the use of attribute information resulted in more precise rule generation.

ROUGE (Recall-Oriented Understudy for Gisting Evaluation) is another metric for evaluating generated text quality. 
ROUGE-1, ROUGE-2 and ROUGE-L scores were computed for both LMN1 and LMN2, where ROUGE-1 captures overlap of single words, ROUGE-2 of two words and ROUGE-L for a sentence of longest common subsequence between the generated rules and the reference texts. Results are shown in Figures \ref{fig:rouge1} and \ref{fig:rouge2}. LMN2 outperforms LMN1 in ROUGE evaluation due to the additional attribute information that allowed for more structured and contextually accurate ABAC rule generation. ROUGE-2 score is relatively lower, specifically for LMN1, as it compares two consecutive words of the generated and the reference text. In LMN1, since attributes are not provided as input, the LLM is using other words related to it or sometimes ignoring that word while doing the comparison.

Processing time comparison was done for both LMN1 and LMN2 to generate the output and the results are shown in Figure \ref{fig:time}. It is observed that LMN1 is faster than LMN2 in most cases. This is primarily because LMN1 uses only one input file, which allows the model to dynamically determine the required attributes, reducing the amount of input data that needs to be processed. LMN2 takes longer since it involves the use of an additional attribute file, which helps the model generate more accurate rules but increases processing time. It also assumes the availability of the desired attribute set as an additional input from the SSO.
Thus, in scenarios where speed is crucial and attribute accuracy can be slightly relaxed, LMN1 is the preferred choice. However, when precision and contextual accuracy are of higher importance, LMN2, despite taking slightly longer, is more suitable.

\begin{figure}[t]
    \centering
    \includegraphics[width=0.6\textwidth]{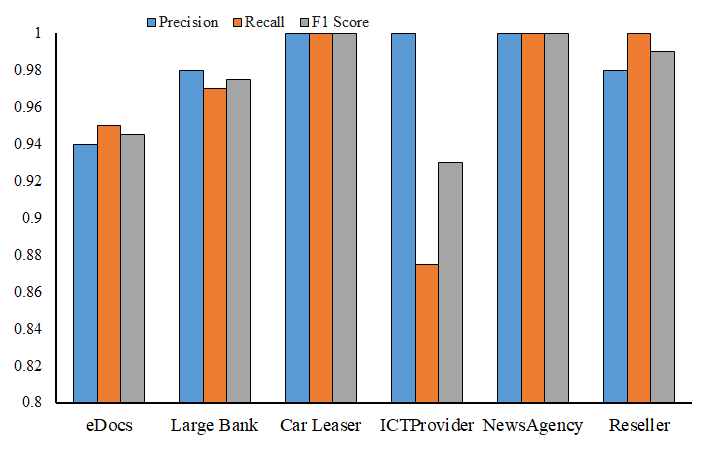}
    \caption{BERTScores for Real Data}
        \label{fig:real_data}
\end{figure}
\subsection{Results on Real Case Study Data}
\label{subsec:realeval}

We also experimented with six real case study data, namely, eDocs, Large Bank, Car Leaser, ICTProvider, NewsAgency  and Reseller as introduced in \cite{Decat2014TheEC}. The BERTScores are plotted in Figure \ref{fig:real_data}. It is observed that all the three metrics, namely, Precision, Recall and F1 Score, have values close to one, depicting high accuracy.

\section{Related Work}
\label{sec:related}

There are broadly two categories of work on generating ABAC rules from existing systems. One of these assumes that the list of authorizations is available in the form of an Access Control Matrix or a set of Role-based Access Control roles and permissions \cite{xu_stoller}\cite{das_et_al}\cite{nakul}. 
The other category, which is relatively a more recent effort, involves deriving ABAC rules from security policies specified in a natural language. 
One of the early attempts towards this was made by Alohaly et al. \cite{AlohalyT019}, which extracts only the attributes from natural language ABAC policies. A Convolutional Neural Network (CNN) is used for this purpose. However, this approach does not produce the complete ABAC policies. In contrast, Xia et al. \cite{9995559} generate ABAC policies from natural language text. But their work is restricted only to the healthcare domain. 
Slankas et al. \cite{Slankas} propose a method for extracting relations among different items from natural language artifacts in text, often penned down by domain experts. It uses type dependency graph for rule extraction. 
In another related work, Masoud et al. \cite{masouddbsec2017} start with tokenization and sentence segmentation, followed by lexical parsing with the help of the CoreNLP toolkit.

Very recently, Abdelgawad et al. \cite{Abdelgawad} have used the spaCy NLP library to obtain ABAC models. A drawback of this approach is the difficulty in obtaining a large set of training data with proper annotation. The associated issues like class imbalance, noisy and incomplete data, etc., further exacerbate the problem. Heaps et al. \cite{userstories} 
use BERT  for access control classification and named entity recognition.  Whether a sentence contains access control information or not is identified by BERT Large transformer. The named entity recognition stage identifies the subjects, objects and the operations that are to be performed in each policy. This is also done by BERT Large transformer. For ensuring correct predictions, attention masks are used to highlight relevant tokens in the input sequence.

None of the existing work presents a web based tool that lets any user upload their NLACP specification and optionally a list of attributes to generate the corresponding machine enforceable ABAC policies harnessing the power of LLMs as done in LMN.

\section{Conclusion}
\label{sec:concl}

We have automated the process of  conversion of Natural Language Access Control Policies into Machine Enforceable Security Policies using Generative Pre-trained Transformer (GPT 3.5). Through careful prompt engineering, we have developed LMN - a freely usable web application that can be used by anyone for doing the conversion. The evaluation metrics, namely, BERTScore, ROUGE and processing time demonstrate the accuracy and efficiency of the tool.
Future work involves supporting different file types like pdf as input, further optimizing the prompts, and using other LLMs.

\section*{Acknowledgment}
The research reported in this paper was partially supported by Cisco Research.

\bibliographystyle{ACM-Reference-Format}
\bibliography{sample-base_LMN}

\clearpage
\newpage
\appendix
\section{Appendix: Prompts used in this work}
\label{apndx_sec_prompts}
\begin{itemize}
    \item \textbf{Prompt 1}
    \begin{itemize}
        \item LMN2:
        \begin{quote}
            Convert the following natural language access control policies into structured ABAC rules using the specified attributes: \texttt{attributes.txt}.\\
            Example Format:\\
            1: (Label: Allow), (Role: User), (Resource: System)\\
            Input:\\
            \texttt{NLACP.txt}
        \end{quote}
        \item LMN1:
        \begin{quote}
            Dynamically extract attributes and generate structured ABAC rules from the following natural language descriptions of access control policies:\\
            Example Format:\\
            1: (Label: Allow), (Role: User), (Resource: System)\\
            Input:\\
            \texttt{NLACP.txt}
        \end{quote}
    \end{itemize}

    \item \textbf{Prompt 2}
    \begin{itemize}
        \item LMN2:
        \begin{quote}
            Here are the natural language access control policies:\\
            \texttt{NLACP.txt}\\
            Convert these using the attributes:\\
            \texttt{attributes.txt}.\\
            Example Format:\\
            1: (Label: Allow), (Role: User), (Resource: System)
        \end{quote}
        \item LMN1:
        \begin{quote}
            Here are the natural language access control policies:\\
            \texttt{NLACP.txt}\\
            Dynamically extract and apply attributes to format them as shown below:\\
            Example Format:\\
            1: (Label: Allow), (Role: User), (Resource: System)
        \end{quote}
    \end{itemize}

    \item \textbf{Prompt 3}
    \begin{itemize}
        \item LMN2:
        \begin{quote}
            Please convert the following descriptions into structured ABAC rules using provided attributes:\\
            Attributes:\\
            \texttt{attributes.txt}\\
            Policies:\\
            \texttt{NLACP.txt}\\
            Example Format:\\
            1: (Label: Allow), (Role: User), (Resource: System)
        \end{quote}
        \item LMN1:
        \begin{quote}
            Please convert the following descriptions into structured ABAC rules by extracting necessary attributes:\\
            Policies:\\
            \texttt{NLACP.txt}\\
            Example Format:\\
            1: (Label: Allow), (Role: User), (Resource: System)
        \end{quote}
    \end{itemize}

    \item \textbf{Prompt 4}
    \begin{itemize}
        \item LMN2:
        \begin{quote}
            Attributes:\\
            \texttt{attributes.txt}\\
            Policies:\\
            \texttt{NLACP.txt}\\
            Format rules with attributes.
        \end{quote}
        \item LMN1:
        \begin{quote}
            Policies:\\
            \texttt{NLACP.txt}\\
            Extract attributes and format rules.
        \end{quote}
    \end{itemize}

    \item \textbf{Prompt 5}
    \begin{itemize}
        \item LMN2:
        \begin{quote}
            Here are some access control policies:\\
            \texttt{NLACP.txt}\\
            Using the attributes provided below, format these policies into structured ABAC rules.\\
            Attributes List:\\
            \texttt{attributes.txt}\\
            Please ensure each rule is clearly defined with all relevant details.
        \end{quote}
        \item LMN1:
        \begin{quote}
            Here are some access control policies:\\
            \texttt{NLACP.txt}\\
            Extract necessary attributes from these descriptions and format them into structured ABAC rules.\\
            Ensure each rule is clear and includes all important elements.
        \end{quote}
    \end{itemize}

    \item \textbf{Prompt 6}
    \begin{itemize}
        \item LMN2:
        \begin{quote}
            Please analyze the following detailed access control policies and use the specified attributes to format these descriptions into structured Attribute-Based Access Control (ABAC) rules.\\
            Given Attributes:\\
            \texttt{attributes.txt}\\
            Policies to be formatted:\\
            \texttt{NLACP.txt}\\
            Each ABAC rule should encapsulate all critical details such as roles, permissions, and conditions applicable to the access control scenario. Ensure the rules are well-defined and actionable in a real-world system security context.
        \end{quote}
        \item LMN1:
        \begin{quote}
            Please dynamically extract critical attributes from the following detailed access control policies and convert them into well-structured Attribute-Based Access Control (ABAC) rules.\\
            Policies to be formatted:\\
            \texttt{NLACP.txt}\\
            Focus on extracting roles, permissions, resources, and any conditions specified within the text. Format these elements into clear and actionable ABAC rules that could be directly implemented in a real-world security system.
        \end{quote}
    \end{itemize}
\end{itemize}
\end{document}